\documentclass[12pt,showpacs,preprintnumbers,amssymb]{revtex4}
\usepackage{epsfig}

\begin{document}

\title{ Critical behaviors in central and peripheral
collisions: a comparative analysis}

\author
{M.~D'Agostino$^{(1)}$, M.~Bruno$^{(1)}$, F.~Gulminelli$^{(2)}$,
R.~Bougault$^{(2)}$, F.~Cannata$^{(1)}$, Ph.~Chomaz$^{(3)}$, 
F.~Gramegna$^{(4)}$, N.~Le~Neindre$^{(3,1)}$, G.~V.~Margagliotti$^{(5)}$,
A.~Moroni$^{(6)}$, G.~Vannini$^{(1)}$, J.~P~.~Wieleczko$^{(3)}$
}

\affiliation
{
(1) Dipartimento di Fisica and INFN, Bologna, Italy \\
(2) LPC Caen (IN2P3-CNRS/ISMRA et Universit\'e),
F-14050 Caen C\'edex, France \\
(3) GANIL (DSM-CEA/IN2P3-CNRS), B.P.5027,F-14021 Caen C\'edex, France\\
(4) INFN Laboratorio Nazionale di Legnaro, Italy\\
(5) Dipartimento di Fisica and INFN, Trieste, Italy\\
(6) Dipartimento di Fisica and INFN, Milano, Italy\\
}

\begin{abstract}
{ \small Quasi-projectile events from peripheral $35 \ A.MeV$ $Au
+ Au$ collisions are compared to central $Au + C$, $Au + Cu$ and
$Au + Au$ events in the same range of excitation energy in terms
of critical partitions and critical exponents. All the different
data sets coherently point to a value $E^*_c=4.5 \ A.MeV$ for the
apparent critical excitation energy. The critical exponents
$\tau, \sigma$ are compatible with the observation of a
liquid-gas phase transition for these systems.}
\end{abstract}


\maketitle

\section{Introduction}

Multifragmentation has been tentatively connected to a critical
phenomenon~\cite{histo_second} since the first heavy ion
experiments. The recent determination of a consistent set of
critical exponents in different multifragmentation
data~\cite{eos,michela} tends to confirm this hypothesis though
finite size corrections to scaling are unknown to a large extent.

Deviations from critical behavior have been however observed for
heavy sources~\cite{fopi,mini100,bill,prlmic95,mfra} formed in
central collisions and the question arises whether the
multifragmentation mechanism depends on the entrance channel.
Indeed thermal scaling near a critical point is expected in the
case of equilibrated sources, which decay independently of the
way they were formed. High energy experiments~\cite{fopi,mini100}
observe partitions strongly governed by the collective energy,
and the deviations of the charge distribution from a power law of
exponent $\tau \approx 2$ can be explained by dynamical
models~\cite{daniel}.

However, an apparent exponent $\tau \ll 2$ has been
observed~\cite{bill,prlmic95,mfra} also in central collisions
measured at intermediate incident energies where collective
components, if any, represent a small correction to the thermal
excitation energy. A tentative explanation of this phenomenon
suggests that Coulomb interaction  would prevent heavy charged
systems to attain statistical equilibration~\cite{das-pratt}.

In the following, by comparing Quasi-Projectile events from
peripheral $35 \ A.MeV$ $Au + Au$ collisions and central
collisions of the asymmetric $Au + Cu$ reactions at 25 and 35
A.MeV incident energy, we will show that in the latter case the
increase of the Coulomb strength with respect to a $Au$ source
leads to asymptotic partitions of early emitted fragments
coexisting with secondary decay products. Once this sequential
emission is backtraced, thermal scaling is observed for several
moments of charge distributions in good agreement with the QP
data.

A multiparametric fit within a Fisher droplet
ansatz~\cite{fisher,moretto_isis} allows to analyze two
additional data samples (central $Au+C$ and $Au+Au$), at
excitation energies far away from the expected critical energy.

Within this ansatz all the reactions consistently point to the
same set of critical exponents and give indication that the
critical excitation energy is $E^*_c=4.5 \pm 0.5 \ A.MeV$. Only
for the heaviest system (central $Au+Au$ at 35 A.MeV) a deviation
from thermal scaling is observed, but more data at lower incident
energy are needed to confirm this point.

The plan of the paper is as follows. After a short presentation
in Section II of the different data samples collected at the NSCL
National Laboratory with the MULTICS-MINIBALL apparatus, the
scaling properties of $Au$ Quasi-Projectiles detected in
peripheral collisions of the $35 \ A.MeV \ Au+Au$ reaction will be
reviewed in Section III. The different central data sets are
presented in Section IV, where the correlation function technique
to backtrace secondary fission is explained, and analyzed in
Section V in terms of Fisher scaling. Conclusions are drawn in
Section VI.

\section{Experiment}

All the measurements discussed in this paper were performed at
the K1200-NSCL Cyclotron of the Michigan State University. The
MULTICS and MINIBALL arrays were coupled to measure charged
products, with a geometric acceptance larger than 87\% of $4\pi$.
More experimental details have been reported in
Ref.~\cite{physlett}.

Peripheral collisions of predominantly binary character were
selected for the reaction $Au+Au$ at 35 A.MeV~\cite{michela}. A
Quasi-Projectile source of nearly constant charge and excitation
energy from 1 to about 8 A.MeV was reconstructed through an
event shape analysis. The fragment angular
distributions~\cite{pal2002} are compatible with an isotropic
emission. The reproduction of the experimental charge partitions
by statistical model calculations~\cite{bondorf} is consistent
with a good degree of source
equilibration~\cite{prctrieste,michela}.

For the reactions $Au+Cu$ at 25 and 35 A.MeV the most central
10\% of the total measured cross section was selected by a charged
particle multiplicity cut. Well detected events (total detected
charge larger than 90\% of the total charge) were analyzed in
terms of event shape and the selected almost spherical events
($\Theta_{flow} \geq 60^o$) are hereafter presented. The same
procedure has been followed to select central events in the
$Au+Au$ reaction, while in the $Au+C$ collision at 25 A.MeV the
average fragment multiplicity $<M>= 2.2 $ (standard deviation
0.2) is too low to perform any shape analysis and only the total
multiplicity and total detected charge cut have been applied.

Mass numbers for clusters of charge 1 and 2 were measured. To
determine the cluster mass distribution, clusters of a given
charge $Z$ were counted on an event by event basis and it was
assumed that $N(A) = N(Z)$. Checks were made on the influence of
the mass estimation. Our main conclusions do not change if we
estimate the cluster mass from the $A_0$-to-$Z_0$ ratio of the
fragmenting system or we assume the mass of stable nuclei.

\section{QP source : Scaling of static observables}

Renormalization group (RG) arguments lead, near the critical point,
size distributions scaling as~\cite{stauffer}

\begin{equation}
 n(A,\epsilon) = q A^{-\tau }f\left( \epsilon A^{\sigma }\right)
\label{scaling}
\end{equation}

where $n(A)=N(A)/A_0$ is the cluster distribution normalized to the
size of the fragmenting system,
$\epsilon$ measures the distance from the critical point
($\epsilon  = (E^*_c-E^*)/E^*_c$ for events sorted in excitation energy
bins), $q$ is a normalization
constant~\cite{stauffer,elliott-scaling}, $f$ is the
scaling function and $\tau,\sigma$ are universal critical exponents.

\begin{figure}[ht]
\begin{center}
\vspace{-1.cm}
\epsfig{figure=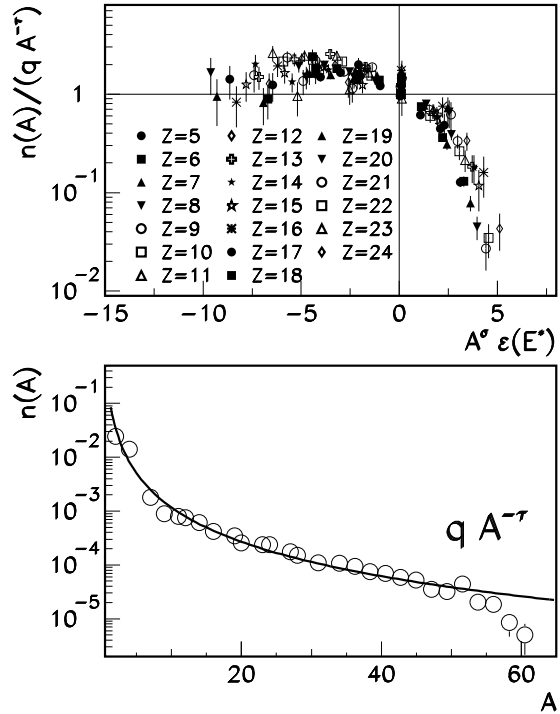,height=12.cm}
\vspace{-1.cm}
\caption {\label{f1}
\it  Peripheral $Au+Au$ collisions.
Top panel: scaling function for the QP source with the ansatz
eq.(\protect{\ref{scaling}}).
Bottom panel: mass distribution of the $QP$ decay products in the
range $4 \le E^* \le 5$ A.MeV,
line: power law with $\tau=2.12$ \protect{~\cite{michela}}.
}
\end{center}
\end{figure}

For peripheral $Au+Au$ collisions the critical exponents as well
as the critical excitation energy of the Quasi-Projectile source
(QP) were extracted from the analysis of the moments of the charge
distribution~\cite{michela}. The critical exponents ($\tau =
2.12, \sigma=0.64$) form a consistent set and are close to the
values of the liquid-gas universality class. Near the excitation
energy identified as critical from the behavior of the biggest
fragment~\cite{michela}, $E^*_c \approx 4.5 \ A.MeV$, the size
distribution is well described by a power law of exponent $\tau =
2.12$ as shown in the bottom panel of Fig.\ref{f1}.

By applying eq.(\ref{scaling}) to the QP events, a scaling
behavior is observed (Fig.\ref{f1} top panel) over the whole range
in excitation energy (from 1 to about 8 A.MeV) and fragment charge
$5 \le Z \le 24$.
Since for the lightest fragments side feeding or
non-equilibrium effects could affect the final production rate,
fragments with $Z < 5$ were discarded.

In recent papers~\cite{elliott-scaling,moretto_isis} Elliott et
al. analyze the critical behavior of the size distribution in
terms of the Fisher droplet model. In this model~\cite{fisher}
the vapor coexisting with a liquid in the mixed phase of a
liquid-gas phase transition is approximated by an ideal gas of
clusters. A scaling around the critical point, similar to
eq.(\ref{scaling}) is assumed, but a different form is suggested
for the scaling function

\begin{equation}
 n(A,T) = q A^{-\tau } exp(\frac{ \Delta G(A,T)}{T})
\label{fisher}
\end{equation}

where $\Delta G(A,T)$ represents the variation of the Gibbs free energy
upon formation of a drop of A nucleons out of an
homogeneous nuclear vapor at temperature $T$

\begin{equation}
\Delta G(A,T) = A \Delta\mu-c_0\epsilon A^\sigma
\label{deltag}
\end{equation}

Here $\Delta\mu$ represents the difference in chemical potential between
the two phases, $\epsilon$ measures the distance from the critical point
$\epsilon = (T_c-T)/T_c$, and  $c_0$ is the
surface energy coefficient.
This expression can be extended to charged systems~\cite{dorso}
by calculating~\cite{bondorf} the Coulomb part of the
free energy $\delta C$ in the Wigner-Seitz approximation

\begin{equation}
\Delta G_{tot}(A,Z,T) =  \Delta G(A,T) - \delta C=
\Delta G(A,T) - \frac{3}{5}e^2\frac{Z^2}{R}
\Bigl (1 - \left ( 1 - c_c \epsilon^\beta \right )^{1/3} \Bigr )
\label{free}
\end{equation}

where $R$ is the radius of a fragment of mass $A$, $c_c$ is a
constant related to the density of the liquid phase~\cite{dorso}
and $\beta$ is a critical exponent.

To verify the ability of this ansatz to extract information about
criticality, we performed a multi-parameter fit of the QP data
(over 200 data points). The values of the critical parameters
from the fit were then compared with those already known from the
analysis of the moments of the charge distribution~\cite{michela}.

As in Ref.~\cite{moretto_isis} the temperature $T$ in
eqs.(\ref{fisher}, \ref{deltag}) has been replaced by $\sqrt{8
E^*/A}$. This ansatz has not to be interpreted as a realistic
temperature estimator, since the assumption that close to the
critical point nuclear systems behave like a degenerate Fermi gas
is highly disputable and not supported by experimental
data~\cite{natowitz}. This Fermi gas ansatz has only to be
considered as an effective parameterization of the scaling
function which allows a direct and quantitative comparison with
the similar analysis performed on Isis data~\cite{moretto_isis}.
In principle $\Delta\mu$ as well as $c_0$~\cite{moretto_isis} and
$c_c$ can depend on temperature. To keep parameters under control
we have taken the simplest parameterizations that allows a good
quality fit of the whole cluster size distribution. $c_0$ and
$c_c$ were treated as free parameters, independent from the
temperature, the exponent $\beta$ was fixed at the value $1/3$ as
resulted from previous analysis on the QP source~\cite{michela}.
Whatever is the parameterization chosen for $\Delta\mu$, its
value at $E^*=E^*_c$ is compatible with zero.

\begin{table}[h!]
{\small
\begin{center}
\begin{tabular}{|c|c|c|c|c|c|}
\hline
{\bf Peripheral collisions}
& $\tau$& $\sigma$& $E^*_c \ (A.MeV)$ & $c_0$ & $\chi^{2}$\\
\hline
$\delta C =0, \Delta\mu$ fixed at 0&
2.05 $\pm$ 0.01 & 0.66 $\pm$ 0.06 & 4.19 $\pm$ 0.06 & 3.3 $\pm$ 0.6 & 1.5\\
\hline
$\delta C =0, \Delta\mu=const.$&
2.07 $\pm$ 0.02 & 0.70 $\pm$ 0.03 & 4.5 $\pm$ 0.4 & 3.0 $\pm$ 0.3 & 1.6\\
\hline
$\delta C = 0$, $\Delta \mu=f(T)$&
2.10 $\pm$ 0.02 & 0.66 $\pm$ 0.02 & 4.5 $\pm$ 0.1 & 8.0 $\pm$ 0.3& 1.3\\
\hline
$\delta C$ eq.(\protect{\ref{free}}), $\Delta \mu=f(T)$&
2.08 $\pm$ 0.02 & 0.66 $\pm$ 0.02 & 4.40 $\pm$ 0.05 & 10.9 $\pm$ 0.4& 1.7\\
\hline
\end{tabular}
\protect\caption{\label{tabparaqp} \it Values of the critical
exponents $\tau$, $\sigma$, critical excitation energy $E^*_c$,
surface energy coefficient $c_0$ and $\chi^2$ of the Fisher
scaling fit through
Eq.s(\protect{\ref{fisher},\ref{deltag},\ref{free}}).}
\end{center}
}
\end{table}

In Table~\ref{tabparaqp} we report the results obtained for two
different parameterizations for $\Delta\mu$ (a constant value or a
polynomial of order two in  $E^*/A$) together with some results obtained
switching on/off the Coulomb term of eq.\ref{free}.
Checks were also made on the parameterization chosen for the Coulomb term.
Results compatible with those reported in
Table~\ref{tabparaqp} are obtained with the parameterization used in
Ref.~\cite{moretto_isis}.

\begin{figure}[ht]
\begin{center}
\vspace{-1.cm}
\epsfig{figure=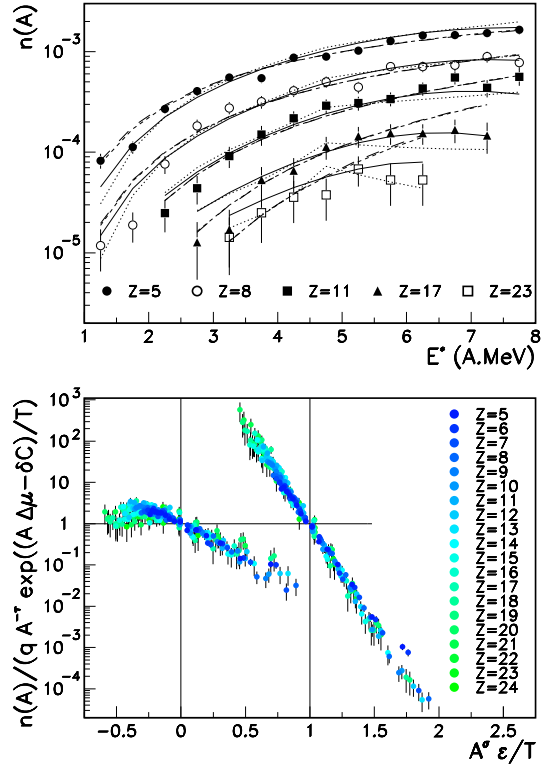,height=12.cm}
\vspace{-1.cm}
\caption {\label{f2} \it  Top panel: Cluster yields of the QP
source as a function of the excitation energy (symbols) and fits
with eq.(\ref{fisher}). Dotted line: $\delta C$
eq.(\protect{\ref{free}}) and $\Delta \mu=f(T)$, solid line:
$\delta C =0$ and $\Delta\mu=f(T)$, short dashed line: $\delta C
=0$, $\Delta\mu=0$, long dashed line: $\delta C =0$,
$\Delta\mu=const.$. Bottom panel: scaled yield distribution
versus the scaled temperature for the cases: $\delta C =0$,
$\Delta\mu=0$ (left) and $\Delta\mu=f(T)$ (right). To visualize
the results on the same picture a constant horizontal shift $C=1$
is given to the latter distribution.
 }
\end{center}
\end{figure}
The fit through Fisher scaling function (eq.\ref{fisher} and
following) produces a scaling behavior for the data, comparable
to the RG ansatz as can be seen in Fig.\ref{f2} (bottom panel).
Critical exponents and the critical excitation energy, reported in
Table~\ref{tabparaqp}, are in agreement with the previous
analyses~\cite{michela} of the moments of the charge distribution.
It is interesting to remark that similarly to the RG ansatz the
scaling obtained with the Fisher scaling function is observed in
the sub-critical and in the supercritical domain, even if in the
context of the Fisher model eq.(\ref{fisher}) has no physical
meaning for $E^*>E^*_c$. This is more quantitatively shown in
Figure \ref{f2} (top panel) which compares the measured
production yields of a few clusters with the results of the fit.
As already observed in Ref.~\cite{fisher_noi} a parameterization
$\Delta\mu = f(T)$ allows the Fisher scaling function to
reproduce the data well above the critical excitation energy.
On the contrary, when $\Delta\mu$ is treated as a constant, the scaling
is observed, mainly for large cluster sizes, only up to the 
critical excitation energy.

The similarity of the results obtained with the RG and with the
Fisher scaling demonstrates that the values of the critical
parameters are very robust with respect to the detailed shape
assumed for the scaling function. On the other hand this also
implies that the adequacy of the Fisher ansatz to the data is not
a sufficient condition to give any detailed physical meaning to
the value of the fit parameters $\Delta \mu$, $c_0$, $c_c$ nor to
the sorting parameter $\sqrt{8 E^*/A}$. As an example, the
quality of the fit and the value of the critical parameters do
not drastically change if one suppresses the Coulomb as well as
the volume term~\cite{elliott-scaling} (see
Table~\ref{tabparaqp}); the value of $c_0$ strongly depends on
the parameterization of the scaling function and the full scale
of the scaled yields depends also on the level density parameter
assumed to estimate the temperature.

\begin{table}[h!]
{\small
\begin{center}
\begin{tabular}{|c|c|c|c|c|c|}
\hline
{\bf Peripheral collisions}
& $\tau$& $\sigma$& $E^*_c \ (A.MeV)$ & $c_0$ & $\chi^{2}$\\
\hline
$E^*= 1-2 \ (A.MeV)$&
2.5$ \pm$ 0.2 & 0.66 $\pm$ 0.03 & 4.5 $\pm$ 0.6 & 10.0 $\pm$ 0.7 & 1.4\\
\hline
$E^*= 2-3 \ (A.MeV)$&
2.1 $\pm$ 0.1 & 0.65 $\pm$ 0.03 & 4.5 $\pm$ 0.3 & 10.0 $\pm$ 0.9 & 1.\\
\hline
$E^*= 3-4 \ (A.MeV)$&
2.10 $\pm$ 0.08 & 0.68 $\pm$ 0.05 & 4.5 $\pm$ 0.2 & 10.0 $\pm$ 1.5& 0.7\\
\hline
$E^*= 4-5 \ (A.MeV)$&
2.08 $\pm$ 0.04 & 0.66 $\pm$ 0.02 & 4.5 $\pm$ 0.2 & 10. $\pm$ 5.& 1.0\\
\hline
$E^*= 5-6 \ (A.MeV)$&
2.08 $\pm$ 0.02 & 0.66 $\pm$ 0.06 & 4.50 $\pm$ 0.02 & 10.1 $\pm$ 0.2& 1.1\\
\hline
$E^*= 6-7 \ (A.MeV)$&
2.08 $\pm$ 0.02 & 0.66 $\pm$ 0.04 & 4.50 $\pm$ 0.02 & 10.1 $\pm$ 0.1& 1.5\\
\hline
$E^* \ge 7 \ (A.MeV)$&
2.08 $\pm$ 0.02 & 0.66 $\pm$ 0.03 & 4.46 $\pm$ 0.02 & 10.0 $\pm$ 0.1& 1.1\\
\hline
\end{tabular}
\protect\caption{\label{tabstepqp}
\it Values of the critical exponents $\tau$, $\sigma$,
critical excitation energy $E^*_c$, surface energy coefficient $c_0$
and $\chi^2$ of the Fisher scaling fit trough
Eq.s(\protect{\ref{fisher},\ref{deltag},\ref{free}}) in bins of
excitation energy.}
\end{center}
}
\end{table}
The Fisher scaling procedure has also been performed on several
intervals of the excitation energy and the results are summarized
in Table~\ref{tabstepqp}. This test allows to validate a Fisher
scaling analysis on central collisions, where the excitation
energy is distributed in a narrow interval.

As can be seen in Table~\ref{tabstepqp} for all the energy
intervals the fitting procedure recognizes the same critical point
even if the critical energy is not contained in the set of data.
The critical exponents are again in agreement with those
previously obtained.

The uncertainty in the critical exponents and in the critical
excitation energy decreases for increasing $E^*$, signaling that
only an approximate description of the critical region can be
provided by events containing information only on the vapor, as
previously found in the study performed in Ref.~\cite{fisher_noi}
on Lattice Gas events. Another signal is given by the value of
$\Delta\mu$ at $E^*_c$.   While this value is compatible with
zero for all the excitation energy bins $E^*>2$ A.MeV, this is
not the case for the first interval of excitation energy, where it results
$\Delta \mu(E^*=E^*_c)=7.2 \pm 0.1$ A.MeV. 
This does not seem merely related to the distance from the critical
excitation energy : indeed ``overcritical'' events 2 A.MeV above
$E^*_c$ recognize the critical point better than events 2 A.MeV
below.

To conclude, we have shown that the Fisher scaling technique is
extremely powerful for the
determination of the critical exponents and the critical excitation
energy. However, the adequacy of the Fisher ansatz to the data is not
sufficient to give a physical meaning to the other fit
parameters.

\section{Central collisions}
\subsection{Equilibration of the emitting sources}

We have shown in the previous Section that a critical-like
behavior is observed around $E^*_c = 4.5 \ A.MeV$ for QP events.
To explore the effect of the entrance channel and/or the size of
the system we will now turn to the analysis of central collisions.

Firstly we show that the considered central events are compatible
with the decay of equilibrated sources and therefore can be good
candidates for an analysis in terms of thermal scaling. The
relevance of the thermostatistical analysis depends indeed on the
approximation at which an equilibrium is realized. As a general
statement, the degree of approximation of an equilibrium is
indicated by the degree of the agreement of data with statistical
models containing the same constraints as the data.

To check the quality of the source selection criteria, a standard procedure
consists in verifying that events are spherically symmetric in momentum
space. 
\begin{figure}[ht]
\vspace{-1.5cm}
\epsfig{figure=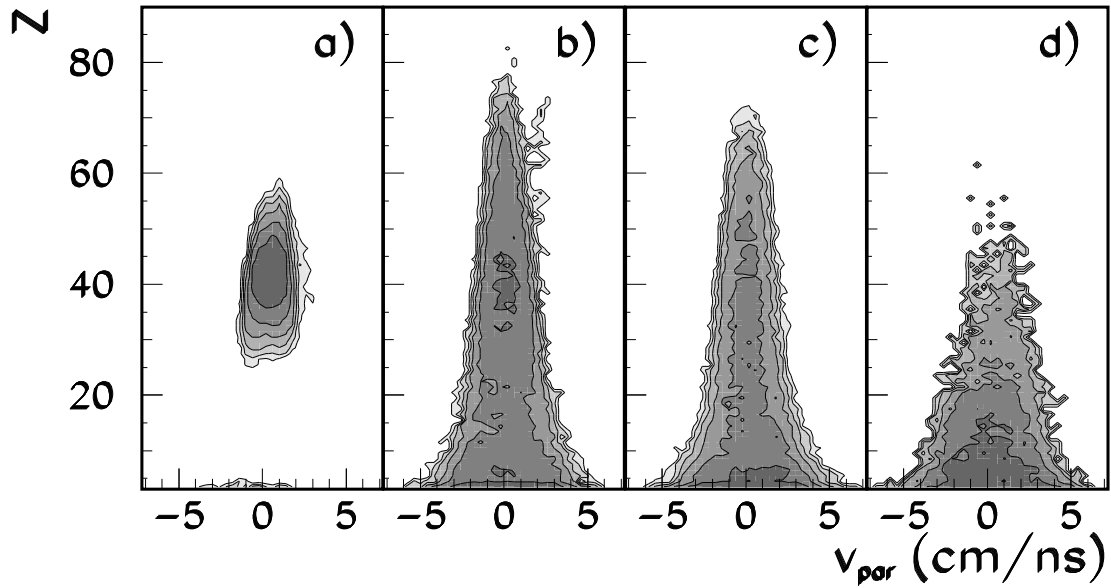,height=11.cm}
\vspace{-1.5cm}
\caption {\label{f6bis} \it  Fragment charge Z versus the
parallel velocity along the beam direction in the center of mass
reference frame for central $Au+C$ events at 25 A.MeV (panel a),
$Au+Cu$ events at 25 (panel b) and 35 A.MeV (panel c) and $Au+Au$
central collisions at 35 A.MeV (panel d). }
\end{figure}
Of course it is very likely that in the dynamical preparation of
the nuclear source the shape degree of freedom may not be completely
relaxed; such a case could in principle be addressed within a statistical
ensemble where deformation is explicitly accounted for. However restricting
the analysis to spherical systems guarantees that preequilibrium
effects do not pollute the statistical sample.

An observable suggested~\cite{lecolley} to discriminate between
single source and multiple sources decay events, is the
correlation between the velocity along the beam direction and the
charge of the fragments.

This observable is shown in Fig.\ref{f6bis} for central
collisions of the reactions $Au+C$ at 25 A.MeV incident energy
(panel a), $Au+Cu$ at 25 (b), 35 A.MeV (c) and $Au+Au$ at 35
A.MeV (d). In this representation one can observe that the source
is at rest in the centre of mass reference frame and also that
the general trend for fragments is to be isotropically emitted.

\begin{figure}[ht]
\begin{center}
\vspace{-2.cm}
\epsfig{figure=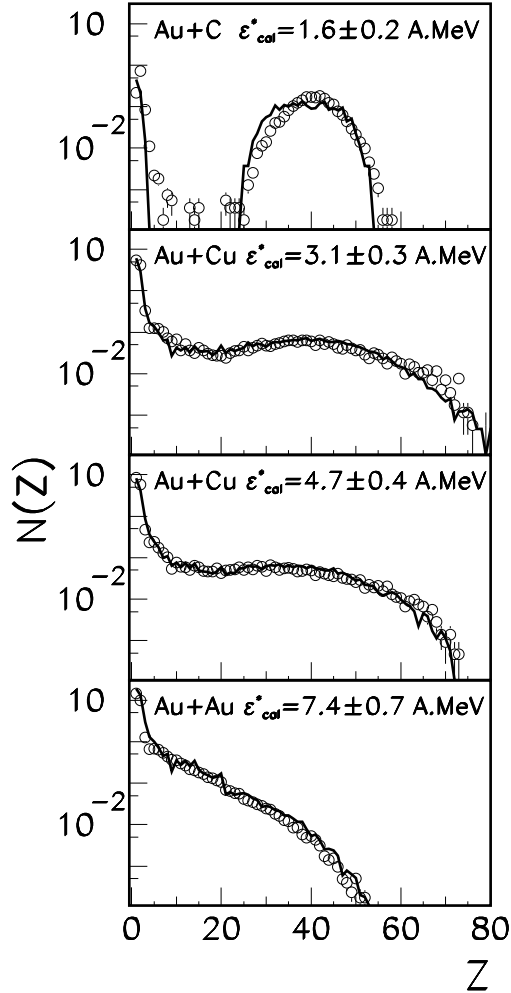,height=16.cm}
\vspace{-1.cm}
\caption {\label{f6}
\it  Symbols: inclusive charge distributions for central $Au+C$ events
at 25 A.MeV, $Au+Cu$ at 25, 35 A.MeV and $Au+Au$ at 35 A.MeV from top
to bottom, respectively.
The reported excitation energy and standard deviation
are the experimental calorimetric values for each sample.
Lines: filtered GEMINI (top panel) and SMM (other panels) simulations.
}
\end{center}
\end{figure}

In Figure \ref{f6} we compare the charge distributions
to the predictions of statistical models.

The $Au+C$ data are the standard prototype of low energy compound
nucleus deexcitation. Due to the strongly forward focused
kinematics and to the minimum detection angle ($\theta_{lab} >
3^o$), complete events are detected only in the case of fission
of the compound. The charge distribution displayed in Figure
\ref{f6} (top panel) is well reproduced by the
Gemini~\cite{charity} code, which describes fragment production
as a sequence of binary fission-like emissions. The excitation
energy needed to the model to reproduce the experimental
distribution (1.4 A.MeV) agrees with the experimental value. The
small shift between the experimental and simulated charge
distributions is presumably due to the (small) difference between
the size of the experimental and simulated source. While Gemini
can not be run for sources heavier than $Au$, in the data the
charge collected by the two fission fragments is the charge of
the completely fused system.

The charge distribution of the central collisions extracted from
the $Au+Cu$ 25 and 35 A.MeV reaction are compared to statistical
SMM model calculations~\cite{bondorf}, where the mass of the
fragmenting source has been assumed equal to the total mass of
the system and the freeze out volume has been taken as $V=3V_0$.
The model well reproduces the data with input excitation energies
(3.3 and 4.1 A.MeV for the 25 and 35 A.MeV data respectively)
close to the average energies measured by calorimetry.

In the case of the $35 \ A.MeV Au+Cu$ sample the excitation energy
is close to the value where the QP shows critical partitions with
$\tau \simeq 2$. However, as it appears from Fig.\ref{f6}, the
central events distribution is much flatter and can be
approximately fitted only with an effective exponent  $\tau_{eff}
\simeq 1.4$. The average excitation energy is lower for the $25 \
A.MeV Au+Cu$ sample, however also for this reaction a two
parameter power law fit of the distribution gives an apparent
exponent $\tau_{eff}\simeq 1.4$.

The decay of the $Au+Au$ source (bottom panel of Fig.\ref{f6} is
also very well described by the statistical SMM
model~\cite{botsep95,pierre}. A moderate contribution of radial
flow (about $1 \ A.MeV$) seems to be superimposed to the thermal
excitation energy and the freeze-out density of this system,
evaluated from the backtracing of experimental data as about
1/3-1/6~\cite{pierre} of the normal density. The apparent
exponent $\tau_{eff}$ of the charge distribution is about $1.2$
and again  the deviation from a power law behavior could be due
in this case to the distance from the critical
point~\cite{raduta} or to some other (size or entrance channel
dependent) physical effects~\cite{milazzoprep}.
\begin{figure}[htb]
\begin{center}
\vspace{-2.cm}
\epsfig{figure=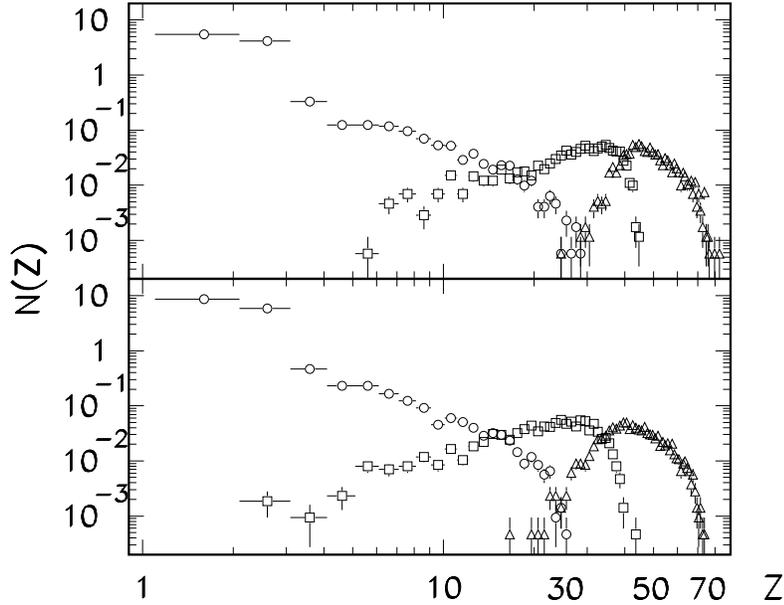,height=12.cm}
\vspace{-1.cm}
\caption {\label{f4}
\it  Charge distribution of the largest fragment (triangles),
the second largest (squares) and the remaining fragments (circles).
Upper (lower) panel: central 25 (35) A.MeV Au+Cu events.
}
\end{center}
\end{figure}

The same kind of agreement with statistical models is also
observed, for all the considered reactions, for the charge
partitions~\cite{botsep95,pierre} as well as for the fragment
kinetic energies. The close similarity between statistical models
and data, together with the isotropy of fragment emission (see
Fig.\ref{f6bis}), suggests that this set of data can be analyzed
an analysis in terms of thermal scaling.

To better understand the low value of $\tau_{eff}$ for the
$Au+Cu$ central collisions, we show in Figure \ref{f4} the charge
distribution of the biggest, second biggest and remaining
fragments for both samples. One can see that the charge
distribution of light fragments (circles) is steeper than the
total distribution and that the flattening, resulting in
$\tau_{eff} \approx 1.4$, is mainly caused by the second largest
fragment, whose distribution covers a wide charge range.

In the next Section we give an interpretation of this behavior.

\subsection{Prompt or sequential emission?}

As suggested in Ref.~\cite{cole}, the observed tail towards small
Z values of the second largest fragment could be due to secondary
decay of the largest primary fragment formed at the freeze-out,
because of the reduced fission barriers of a source heavier than
the $QP$ one.  According to this interpretation, fragments
generated later than the freeze-out time should keep memory of
their later emission by showing a strong velocity correlation.
This should be apparent in the two-fragment velocity correlation
function, defined as

\begin{equation}
1+ R(v_{red}) = C \frac{Y(v_{red})} {Y_{back}(v_{red})}
\end{equation}

where
$v_{red} = ( \vec v_i - \vec v_j  )/ \sqrt {(Z_{i}+Z_j)}$
is the reduced relative velocity of fragments i and j
with charges $Z_i$ and $Z_j$; $Y(v_{red})$ and $Y_{back}(v_{red})$
are the coincidence and background yields for fragment pairs of reduced
velocity $v_{red}$ and $C = N_{back}/N_{coinc}$ where $N_{coinc}$ and
$N_{back}$ are the total number of coincidence and background pairs. The
background yield was constructed by means of the mixed event
technique~\cite{trock}.

\begin{figure}[ht]
\begin{center}
\vspace{-4.cm}
\epsfig{figure=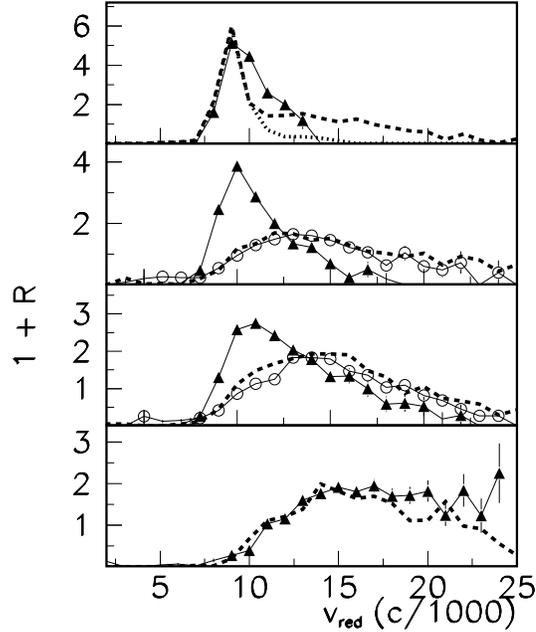,height=15.cm}
\vspace{-1.cm}
\caption {\label{f6-1} \it Correlation functions of the reduced
velocity for the two largest fragments in each event. From top to
bottom: central $Au+C$ events at 25 A.MeV, $Au+Cu$ at 25, 35
A.MeV and $Au+Au$ at 35 A.MeV. Triangles: correlation between
detected fragments, open symbols: the same correlation after
re-clusterization for central events(see text). Dashed lines: QP
correlation at the same excitation energy as the corresponding
central events. }
\end{center}
\end{figure}
This correlation function is in principle affected by a space-time
ambiguity. However, if the density of the sources is comparable
(as indicated by statistical model predictions) this observable
measures the fragment formation time.

Figure \ref{f6-1} displays the correlation functions of the
reduced velocity for all the sets of central events. These
observables  are also compared to the correlations obtained from
$QP$ events at the same excitation energy as the central sets.

For central $Au+C$ events at 25 A.MeV the correlation function is
very similar to the QP one, especially when only fission events
are considered (dotted line). The peak at $v_{red} \approx 10$
(c/1000)  corresponds to the Coulomb reduced velocity of two
touching charged spheres.

For central $Au + Cu$ events the correlation function of the
reduced velocity between the two heaviest fragments in each event
signals emission times longer than in the QP decay at the same
excitation energy. The Coulomb hole is narrower than in the $QP$
case and the peak corresponding to the reduced velocity of two
touching charged spheres, observed in the $Au+C$ reaction, is
still present even if the excitation energy of the source is
larger.

For the $Au+Au$ central collisions the correlation function of the
reduced velocity (Fig.\ref{f6-1} bottom panel) of the two
heaviest fragments does not give indications of a late emission.

Several interpretation of this behaviour can be given. One is
that for $Au+Au$ central events all the detected fragments were
originated at the freeze-out time. Another alternative
interpretation is that the increased Coulomb strength with
respect to a $Au$ source also affects the decay of fragments
other than the largest one.

More sophisticated correlation techniques would be needed to
answer this question.

For the $Au+Cu$ reactions we can use the correlation function
information to backtrace the sequential emission.
We reclusterize the two heaviest fragments with a Gaussian shaped probability
distribution, centered at the Coulomb reduced velocity and with a width
given by the left tail of the correlation function.
When the two heaviest fragments are reclusterized, they are replaced
in the event by a primary fragment with a charge equal to the sum of their
charges and a velocity vector given by the weighted mean of their
velocities.
As shown in Figure \ref{f6-1}, after this procedure the correlation
function has a very similar shape to the QP one in the same excitation
energy range.

\subsection{Critical exponents}

\begin{figure}[ht]
\begin{center}
\vspace{-2.cm}
\epsfig{figure=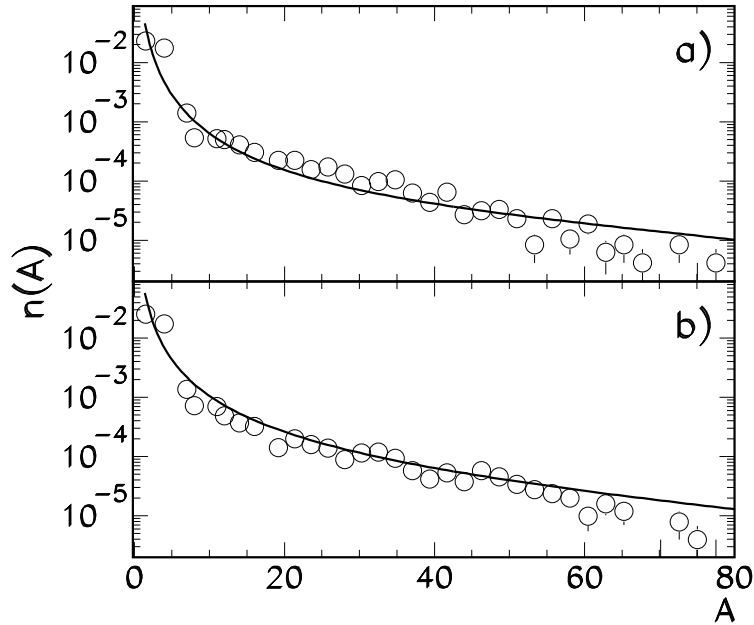,height=12.cm}
\vspace{-1.cm}
\caption {\label{f5-1} \it Cluster size distribution for central
$Au+Cu$ events (open symbols), after re-clusterization of the two
heaviest fragments with the procedure described in the text. The
line represents the power-law from
Eq.s(\protect{\ref{fisher},\ref{deltag},\ref{free}}) describing
the QP events at the same excitation energy as central $Au+Cu$ 25
A.MeV collisions (top panel) and $Au+Cu$ 35 A.MeV collisions
(bottom panel). }
\end{center}
\end{figure}

After the backtracing of secondary fission the cluster size
distribution for the central collisions $Au+Cu$ at 25 and 35 A.MeV
are in very good agreement with the QP size distributions at the same
excitation energy (see Fig.\ref{f5-1}).
In particular the 35 A.MeV reaction follows a power-law with
$\tau \approx 2$ as appears from the bottom panel of
Fig.\ref{f5-1}. The fact that at similar excitation energies the
central and peripheral sources show a similar fragment
distribution strongly supports the thermal scaling.

In addition this means that, at least up to a source charge around
100, the increase of the Coulomb in the entrance channel only
affects the decay of primary fragments, but not the bulk
multifragmentation.

\begin{figure}[ht]
\begin{center}
\vspace{-2.cm}
\epsfig{figure=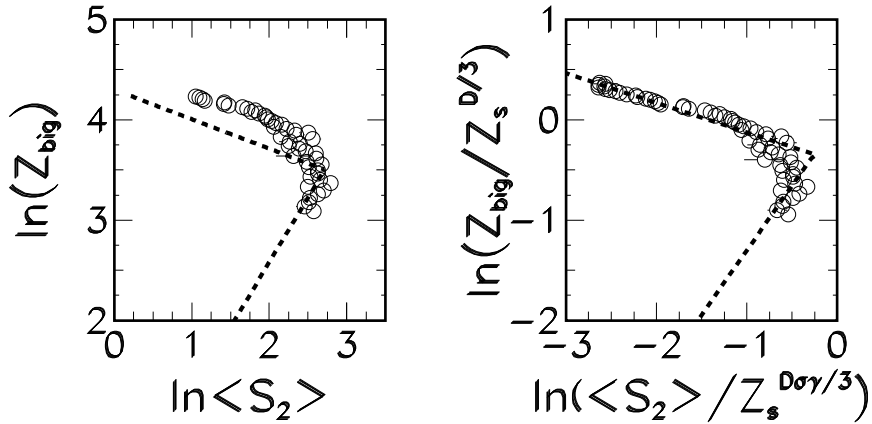,height=8.cm}
\vspace{-1.cm}
\caption {\label{f5-2}
\it  Campi plot for the QP source (dashed line)
and for $Au+Cu$ 35 A.MeV central collisions (symbols). In the left
panel the unscaled distribution is given, in the right panel scaled
moments are shown.
}
\end{center}
\end{figure}

A power-law with exponent $\tau \approx 2$ corresponds to
partitions with very large size fluctuations~\cite{cdorso}, as
signaled by the Campi plot of the central events
(Fig.\ref{f5-2}), showing a ``liquid'' as well a ``gas'' branch.

It has to be noted, however, that a bimodal shape of the Campi plot is a
necessary, but not sufficient, signal to claim the observation of a
phase transition. Indeed simulations, which do not contain
criticality, but only take into account the phase
space~\cite{elliott-scaling} can produce Campi-plots apparently
similar in shape. Only critical exponents~\cite{rassegna}
allow to identify criticality.

A strong check of compatibility among QP critical events and
$Au+Cu$ central events consists in the finite size scaling of the
moments of the size distribution.

In Ref.~\cite{jaq91} it has been shown that, as for percolation
events, in finite nuclear systems near the critical excitation
energy the size of the largest cluster and the second moment
scale for the size of the system following the hyper-scaling
relation~\cite{jaq91,elliott-scaling}.

Following Ref.~\cite{jaq91} the largest cluster size is expected
to scale as $S^{D/3}$, where $S$ is the size of the system,
$D=\frac{d}{\tau -1}$ is the fractal dimension, $d = 3$ is the
Euclidean dimension. The second moment $S_2$ is expected to scale
as $S^{D \sigma \gamma /3}$.

In the right panel of Fig.\ref{f5-2} the correlation between the
size of the largest fragment and the average second moment is
shown for the central source, scaled~\cite{jaq91} via the
critical exponents of the QP source~\cite{michela}. The observed
agreement shows again the validity of the scaling~\cite{ther_sca}
and the compatibility of critical exponents for the two systems.

In this section we have presented ``compatibility'' checks between the
static observables of emitting sources, different in size and in the
way they were formed.
However, our aim is to extract a complete set of critical exponents
also for sources formed in central collisions.

The procedure adopted for the QP in Ref.~\cite{michela} cannot be
applied to central collisions, since the excitation energy is
distributed in a very narrow interval. However we have shown in
Section III that even for sets of data that do not contain the
critical point it is possible to get information on critical
excitation energy and critical exponents by working with the
Fisher scaling function eq.(\ref{fisher}).

\section{Fisher scaling for central collisions}

In order to get critical parameters for the central events,
independently of the information extracted from the QP data, the
fitting procedure described in Sect.III has been applied to the
four sets of central events discussed above.

\begin{figure}[ht]
\begin{center}
\vspace{-2.cm}
\epsfig{figure=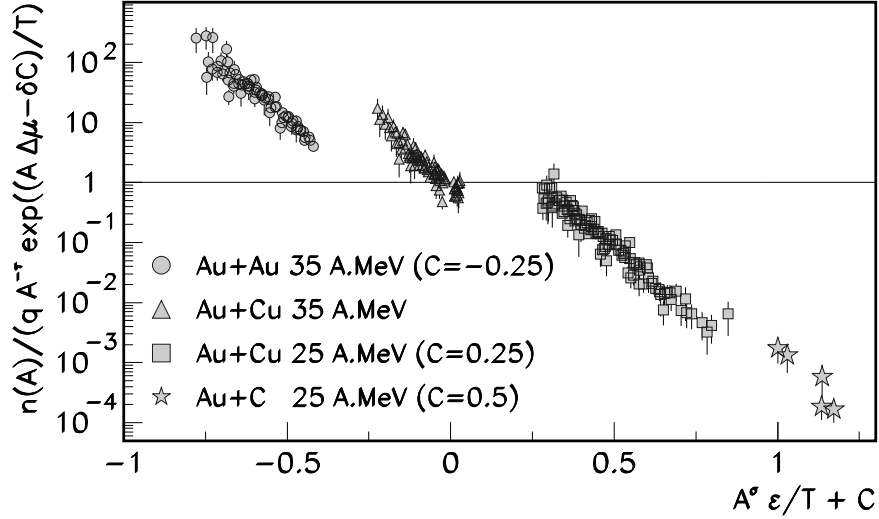,height=10.cm}
\vspace{-1.cm}
\caption {\label{f7} \it  Fit of central events with the scaling
function ansatz eq.(\ref{fisher}). To visualize the results on
the same picture a constant horizontal shift $C$ is given to the
distributions.
 }
\end{center}
\end{figure}

\begin{table}[h!]
{\small
\begin{center}
\begin{tabular}{|c|c|c|c|c|c|}
\hline
{\bf Central collisions}
& $\tau$& $\sigma$& $E^*_c \ (A.MeV)$ & $c_0$ & $\chi^{2}$\\
\hline
Au+C 25 A.MeV&2.0 $\pm$ 0.6&0.64 $\pm$ 0.02&4.5 $\pm$ 0.1&10.0 $\pm$ 0.2&7.8\\
\hline
Au+Cu 25 A.MeV&2.10 $\pm$ 0.04&0.60 $\pm$ 0.02&4.4 $\pm$ 0.1&10.0 $\pm$ 0.6
&1.8\\
\hline
Au+Cu 35 A.MeV&2.05 $\pm$ 0.04&0.68 $\pm$ 0.03&4.5 $\pm$ 0.2&9. $\pm$ 1.&1.3\\
\hline
Au+Au 35 A.MeV&2.1 $\pm$ 0.1&0.68 $\pm$ 0.03&4.51 $\pm$ 0.01&10.0 $\pm$ 0.1
&1.7\\
\hline
All central&2.05 $\pm$ 0.02&0.66 $\pm$ 0.02&4.40 $\pm$ 0.03&13.0 $\pm$ 0.4
&2.0\\
\hline
\end{tabular}
\protect\caption{\label{tabparacen}
\it Values of the critical exponents $\tau$, $\sigma$,
critical excitation energy $E^*_c$, surface energy coefficient $c_0$
and $\chi^2$ of the Fisher scaling fit trough
Eq.s(\protect{\ref{fisher},\ref{deltag},\ref{free}}).
}
\end{center}
}
\end{table}
The resulting scaling functions are shown in Figure \ref{f7} and
the corresponding critical parameters are  reported in
Table~\ref{tabparacen}.

Also the data samples not covering the critical region point to
the same critical energy and to compatible critical exponents. The
fit is of high quality and the values of $\Delta\mu$ at $E^*_c$
are compatible with zero, within the error, except for the
compound nucleus de-excitation events ($Au+C$ reaction). In this
case the value of $\Delta\mu$ at $E^*_c$ is large ($2.9 \pm 0.06$
A.MeV) and the quality of the fit is poor. For this reaction,
however, the excitation energy is very far from the critical one.
As found for the QP events (see Section III) and in the study
performed on a Lattice Gas~\cite{fisher_noi}, an approximate
description of the critical region can be provided by events
containing only information on the vapor.

\begin{figure}[ht]
\begin{center}
\vspace{-2.cm}
\epsfig{figure=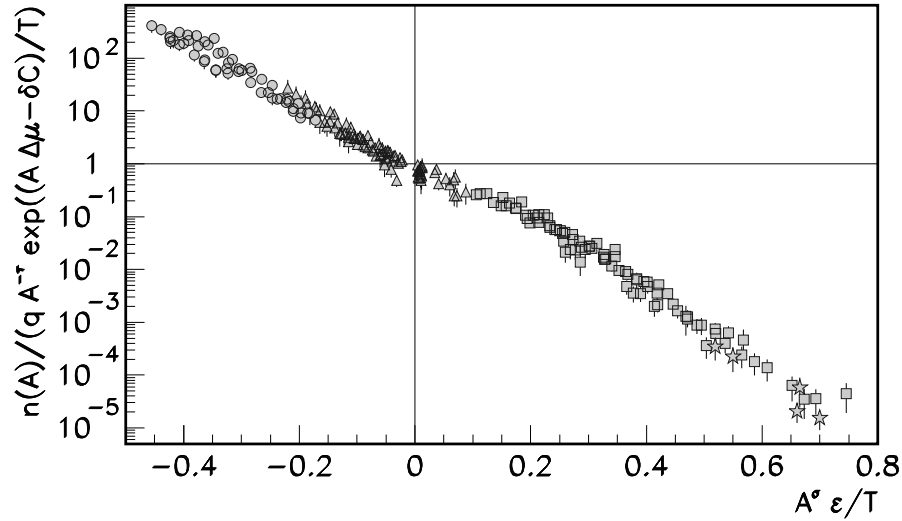,height=10.cm}
\vspace{-1.cm}
\caption {\label{f8}
\it  Simultaneous fit of an equal statistics subsample of
the central events with the scaling function
ansatz eq.(\ref{fisher}). Symbols as in Fig.\protect{\ref{f7}}.
 }
\end{center}
\end{figure}

To check whether the scaling function (\ref{fisher}) is able to
describe the decay of finite charged nuclear matter, independently on
the size of the source, a subset of events of
equal statistics has been selected
for the four sets of central events (1000 events for each reaction)
and Fisher scaling (\ref{fisher}) has been repeated on this sample.
The resulting scaling function is displayed in Fig.\ref{f8} and the
parameters are again reported in Table~\ref{tabparacen} (label : 'All central').
The good quality of the scaling and the perfect agreement of the
critical parameters with the Quasi-Projectile results are a striking
proof of the universality of the scaling: the critical behavior and
the whole shape of the scaling function are independent both on the
size of the system and  on the entrance channel.
This implies that the Fisher scaling technique is extremely powerful
and the determination of the critical parameters is robust with respect to
the possible effects induced by the Coulomb interaction and by
the reaction mechanism.

A word of caution is however necessary. In Section II we have
presented a comparable quality scaling obtained with the RG
ansatz (\ref{scaling}) which applies to generic critical
phenomena like percolation, not necessarily connected to a
thermal transition. It is also well known that the liquid-gas
critical exponents are very close to the percolation ones.
Therefore other non static observables have to be analyzed before
one can identify the transition as belonging to the liquid-gas
universality class, but this is beyond the aim of this
paper~\cite{inprep}.

\section{Conclusions}
The decay of the Quasi-Projectile formed in peripheral 35 A.MeV
$Au + Au$ collisions has been compared to different sets of central
events in terms of critical partitions and critical exponents.

All reactions are compatible with a critical behavior at an
excitation energy of about 4.5 A.MeV, with  critical exponents for
the size distribution $\tau \approx 2, \sigma \approx 0.65$.
Analyzing the velocity correlations between the two heaviest
fragments we have demonstrated that fission is an important
deexcitation channel for the secondary decay of heavily charged
systems ($Z_{source} \approx 100$). If this decay is back-traced
we observe a perfect scaling with the size of the source between
the peripheral QP events ($Z_{source} \approx 80$) and the
central $Au+Cu$ ones ($Z_{source} \approx 100$). On the other
hand the size distribution of central $Au+Au$ events is flatter
than the QP one at a compatible deposited energy, and this fact
cannot be ascribed to secondary fission.

The impressive quality of the scaling and the compatibility of the
results of central and peripheral collisions does not demonstrate
unambiguously the thermal nature of the transition, though it
certainly is a strong circumstantial evidence for its existence.
Indeed the shape of the scaling function (i.e. the parameters
$c_0$ and $\Delta \mu$) depends non negligibly on the technical
assumptions made in the procedure of the fit and a similar scaling
is obtained with the ansatz Eq.(\ref{scaling}) that does not
assume a thermal nature for the transition. This means that the
different size distribution (implying a different behavior for
$\Delta \mu$) observed for the $Au+Au$ system can be ascribed to
a different thermodynamical path for the state variables explored
in central collisions of symmetric systems, but also to non
thermal effects due either to the entrance channel (signaled by
the onset of collective flow~\cite{fopi}) or to the high charge
of the system (Coulomb explosion for too heavily charged
sources~\cite{bill,eos}).

The scaling ``per se'' does not determine the order of the
transition either. Indeed in many different statistical
calculations~\cite{prl99,raduta,richert_last,fisher_noi} first
order phase transitions in finite systems lead to observations
typical of continuous transitions including critical behaviors and
critical exponents. Further work to assess this point is in
progress~\cite{inprep}.

\vskip 0.3cm {\small The authors would like to acknowledge the
{\it Multics-Miniball} collaboration, which has performed the
experiments. The authors are also grateful to P.~F.~Mastinu and
P.~M.~Milazzo for a critical reading of the manuscript.
\\
This work has been partially supported by NATO grants CLG-976861
and by grants of the Italian Ministry of Instruction, University
and Research (contract COFIN99). }

\end{document}